\documentclass[12pt]{article} 
 \usepackage{fullpage}  

\usepackage{graphicx}
\usepackage{dcolumn}
\usepackage{bm}
\usepackage{xcolor} 
\usepackage{textcomp} 
\usepackage[export]{adjustbox} 
\usepackage[english]{babel}
\usepackage{authblk}
\usepackage{breqn}

\begin{document}
\title{Efficient orbital imaging based on ultrafast momentum microscopy and sparsity-driven phase retrieval} 

%

 \author[1,*]{G~S~M~Jansen} 
 \author[1]{M~Keunecke} 
 \author[1]{M~D\"uvel} 
 \author[1]{C~M\"oller} 
 \author[1]{D~Schmitt}
 \author[1]{W~Bennecke}
 \author[1]{F~J~S~Kappert}
 \author[1]{D~Steil}
 \author[2]{D~R~Luke}
 \author[1]{S~Steil}
 \author[1]{S~Mathias}

\affil[1]{1\textsuperscript{st} Physical Institute, University of G\"ottingen, Friedrich-Hund-Platz 1, 37077, G\"ottingen, Germany}
\affil[2]{Institute for Numerical and Applied Mathematics, University of G\"ottingen, Lotzestrasse 16-18, 37083 G\"ottingen , Germany}
\affil[*]{gsmjansen@uni-goettingen.de}
\maketitle
\date{}
\setcounter{Maxaffil}{0}
\renewcommand\Affilfont{\itshape\small}

\begin{abstract}
{
We present energy-resolved photoelectron momentum maps for orbital tomography that have been collected with a novel and efficient time-of-flight momentum microscopy setup. This setup is combined with a 0.5~MHz table-top femtosecond extreme-ultraviolet light source, which enables unprecedented speed in data collection and paves the way towards time-resolved orbital imaging experiments in the future. Moreover, we take a significant step forward in the data analysis procedure for orbital imaging, and present a sparsity-driven approach to the required phase retrieval problem, which uses only the number of non-zero pixels in the orbital. Here, no knowledge of the object support is required, and the sparsity number can easily be determined from the measured data. Used in the relaxed averaged alternating reflections algorithm, this sparsity constraint enables fast and reliable phase retrieval for our experimental as well as noise-free and noisy simulated {photoelectron momentum map} data.
}

\end{abstract}


\section{Introduction}
Angle-resolved photoemission spectroscopy (ARPES) is a powerful method to study the structure \cite{puschnig_reconstruction_2009, luftner_imaging_2014, weis_exploring_2015, willenbockel_energy_2013, kliuiev_application_2016} and electron dynamics \cite{tautz_structure_2007, schwalb_electron_2008, hagen_electronic_2010, varene_ultrafast_2012,  steil_spin-dependent_2013, marks_formation_2014,  schonauer_charge_2016} of adsorbed molecules on metal surfaces. Under suitable conditions, ARPES measures the intensity of the spatial Fourier transform of the initial electronic state $\psi$, multiplied by a polarization-dependent factor \cite{puschnig_reconstruction_2009}. 
The measured signal can then be expressed as 
\begin{equation}
    I(\vec{k}) = |\vec{A}\cdot \vec{k}|^2~|FT\{\psi(\vec{r})\}|^2~\delta(E_b + E_{kin} + \Phi - \hbar \omega),
    \label{eq:basic_arpes}
\end{equation}
where $\vec{A}$ is the vector potential of the ionizing radiation and $\vec{k}$ is the photoelectron momentum. The $\delta$ function ensures energy conservation, taking into account the binding energy $E_b$ of the initial state, the final state kinetic energy $E_{kin}$, the sample work function $\Phi$ and the photon energy $\hbar \omega$.
This Fourier-transform relationship between measurement and object occurs similarly in coherent diffractive imaging (CDI) \cite{miao_extending_1999}, where the intensity of a Fraunhofer diffraction pattern is measured. In CDI, iterative phase retrieval algorithms are then used to reconstruct a high-resolution image of the diffracting object \cite{fienup_reconstruction_1987, marchesini_invited_2007}. This analogy has lead to the development of orbital imaging: a technique in which molecular orbitals are reconstructed from ARPES data using iterative phase retrieval algorithms \cite{luftner_imaging_2014, kliuiev_application_2016}. {Future progress in orbital imaging, however, relies on two important factors: (I) the development of tailored experimental tools for data acquisition, and (II) the development of the necessary mathematical tools in the data analysis procedure.} 

{
For efficient orbital imaging, one first needs to be able to collect energy- and momentum-resolved photoelectron data in an efficient manner. So far, using typical two-dimensional angle-resolved photoelectron spectroscopy (ARPES) detectors, which only have access to a single momentum direction, a measurement of the the full 2D photoelectron momentum map required tedious scanning of the azimuthal angle of the sample. Nowadays, this issue can be overcome by the usage of so-called momentum microscopes, which collect the full azimuthal angle of the photoemitted electrons in one measurement. Moreover, using time-of-flight-based momentum microscopy, scanning of the kinetic energy can be omitted, too. However, using a time-of-flight technique, a pulsed extreme-ultraviolet (EUV) light source is needed for such measurements, which has so far limited the usage of these microscopes to synchrotron radiation facilities. Here, in contrast, we present first orbital imaging data that has been collected with a time-of-flight momentum microscope and a high-repetitive 0.5~MHz femtosecond table-top EUV light source. We show that such a setup is ideally suited for efficient momentum microscopy and further opens the door towards time-resolved orbital imaging experiments in the future.}

In terms of data analysis, efficient phase retrieval algorithms are required, because, like CDI, photoemission measurements do not record the phase of the photoemission pattern, and the phase must be reconstructed in order to achieve an image. In the context of CDI, it has been established that this is possible for objects with a known finite support \cite{fienup_reconstruction_1987} of which the diffraction pattern is sufficiently oversampled \cite{sayre_implications_1952}. Phase retrieval is performed using an iterative algorithm which optimizes the phase such that it best matches the support. A commonly used algorithm is \emph{alternating projections} (also known as \emph{error reduction}), which functions by sequentially applying the support constraint and the measurement constraint. Furthermore, a wide range of projection-based algorithms has been developed to enable efficient and accurate phase retrieval \cite{marchesini_invited_2007}. A common problem in both CDI and ARPES-based orbital imaging is, however, that the object support is unknown. While it has been shown that phase retrieval with a loose support is possible in specific cases (such as for a non-negative object \cite{fienup_reconstruction_1982}), in general a tight support is required \cite{fienup_reconstruction_1987}. In this case, the object and support can be reconstructed simultaneously using a \emph{shrink-wrap} routine in combination with the phase retrieval algorithm \cite{marchesini_x-ray_2003}. This approach has been successfully applied to orbital imaging \cite{kliuiev_algorithms_2018}. Nevertheless, the shrink-wrap procedure leads to a more complicated algorithm which may require manual fine tuning, such as in the determination of the object support guess (as explained well in Ref.~\cite{kliuiev_algorithms_2018}). 
Aside from the support projection, there exist several other projections which may enable phase retrieval \cite{elser_phase_2003}, such as realness, non-negativity or atomicity (where the object consists of multiple identical elements). Such projections may eliminate the need to accurately determine the object support, which simplifies the reconstruction procedure.
Here, we {implement} a sparsity-driven phase retrieval algorithm which does not rely on a support constraint. The sparsity projection \cite{jaganathan_sparse_2013, shechtman_gespar:_2014, yokoyama_sparse_2019}, which keeps the first $S$ brightest pixels of an object and reduces the rest to zero, is easy to implement and, although slower than the support projection, can be calculated efficiently. We demonstrate that the sparsity-driven algorithm enables accurate and efficient reconstruction of typical molecular orbitals from our momentum microscopy measurements, even in the presence of severe Poisson noise. 

\section{The experimental setup}


\begin{figure*}[htbp!]
\includegraphics[width=\textwidth,right]{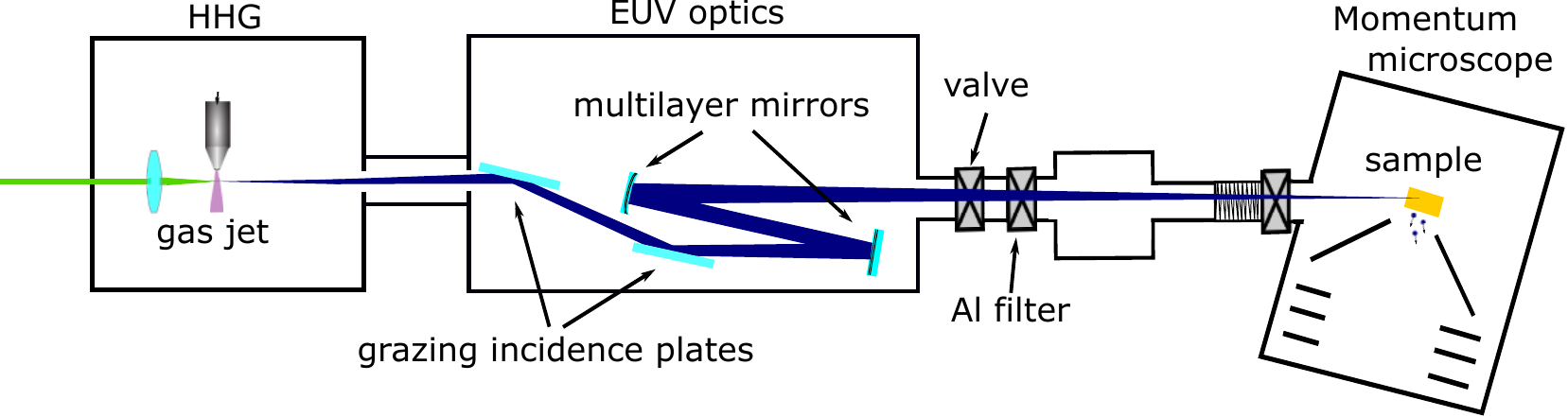}
\caption{
\label{fig:setup}
Schematic overview of the momentum microscopy setup, with the driving laser (40~fs, 515~nm center wavelength, up to 15~\textmu J at  0.5~MHz). A combination of grazing incidence plates (anti-reflection-coated for the driving laser), EUV multilayer mirrors and an 200~nm Aluminum filter is used to select the eleventh harmonic with 26.5~eV photon energy. The second multilayer mirror focuses the EUV light onto the sample, where the momentum and energy of all photoelectrons is measured using the momentum microscope.
}
\end{figure*} 

{
A schematic overview of the EUV beamline used for photoemission is shown in Fig.~\ref{fig:setup}. In this system, 40~fs, 515~nm laser pulses (Active Fiber Systems, 7.5~W at 0.5~MHz repetition rate) are used for high harmonic generation in an argon gas jet. A pair of multilayer mirrors is used to filter out the eleventh harmonic, which has roughly 26.5~eV photon energy, a bandwidth of 140~meV and a pulse length of 20~fs. The second multilayer mirror is curved to focus the high harmonics to a {200~\textmu m}  spot on the sample with an incidence angle of 68\textdegree. Angle-resolved photoemission spectra are collected using a time-of-flight-based momentum microscope (Surface Concept), which enables measurement of the full electron momentum as well as the kinetic energy of all photoemitted electrons \cite{medjanik_direct_2017}. For the settings used to collect the data discussed here, the momentum microscope´s energy and momentum resolution in the vicinity of the molecular orbitals are 70 - 80~meV and 0.04~\AA\textsuperscript{-1}, respectively. The high harmonic generation light source was optimized to yield on average 0.8 detected electrons per pulse in order to avoid space-charge effects \cite{passlack_space_2006, hellmann_vacuum_2009, schonhense_multidimensional_2018}. At this detection rate (0.5 MHz * 0.8 = 0.4 MHz), total integration times for the presented data were 2 hours. An in-detail description of the femtosecond time-resolved momentum-microscopy setup is given in Ref.~\cite{keunecke_time-resolved_2020}, which describes the setup in a state optimized for operation at 1~MHz.}

\section{Sparsity-based orbital reconstruction}
Generally, molecular orbitals are three-dimensional objects. Complete reconstruction therefore requires measurement of a three-dimensional photoemission spectrum and subsequent phase retrieval \cite{weis_exploring_2015}. The three-dimensional spectrum can be built up by performing ARPES while scanning the photon energy. This time-consuming procedure can be avoided for planar orbitals of the form $\psi(x,y)\psi(z)$, where $x$, $y$ and $z$ are the spatial coordinates. The Fourier transform of such orbitals can be expressed as $\hat{\psi}(k_x,k_y)\hat{\psi}(k_z)$, where $k_x$,$k_y$, and $k_z$ are the corresponding momentum coordinates. If the intensity profile $|\hat{\psi}(k_z)|^2$ is known, it is then possible to calculate $|\hat{\psi}(k_x,k_y)|$ from the measured photoemission signal at constant binding energy. From this two-dimensional (2D) photoelectron momentum map, a 2D image of the orbital can then be reconstructed. For sufficiently thin molecules, $|\hat{\psi}(k_z)|^2$ can be assumed to be unity. This approximation works well for the orbitals of planar aromatic molecules at low binding energy, and 2D orbital imaging has been demonstrated in for example pentacene \cite{puschnig_reconstruction_2009, luftner_imaging_2014, kliuiev_application_2016, kliuiev_algorithms_2018}, sexiphenyl \cite{puschnig_reconstruction_2009} and PTCDA \cite{luftner_imaging_2014}.

Initially {the support of the measured orbital, i.e. the set of pixels for which its value is nonzero,} is unknown in orbital imaging. Nevertheless, the rough size of the orbital can usually be estimated, for example from the intermolecular distance measured by low energy electron diffraction. This yields the number of non-zero pixels $S$ in the orbital image without imposing a specific support. Furthermore, the molecular orbital can usually be described by a compact arrangement of positive and negative lobes. Based on these considerations, we propose a combined sparsity and realness projection $P_{S,real}$ operating on the current orbital guess $\psi_n$. The projection $P_{S,real}(\psi_n$) is calculated as follows: 
\begin{enumerate}
    \item Set the imaginary part of the guess to zero, leaving the real part.
    \item Sort the guess by its amplitudes $|\psi_n|$ and find the amplitude $a$ of the S\textsuperscript{th}-most bright pixel.
    \item Of the unsorted guess $\psi_n$, set all pixels to zero where $|\psi_n|<a$. 
\end{enumerate}
In addition to $P_{S,real}$, we use the measurement constraint $P_M$. This means we calculate the predicted momentum map using a Fourier transform, set the amplitudes of the predicted field equal to the measured values and perform an inverse Fourier transform to calculate the updated orbital guess. A single step of the alternating projections (AP) algorithm is then defined as 
\begin{equation}
    \psi_{n+1} = P_{S,real}(P_M(\psi_n)).
\end{equation}
In order to monitor the progress of the reconstruction, we calculate the normalized root mean square (RMS) change between subsequent iterations $\psi_{n}$ and $\psi_{n-1}$. This provides a measure of how well the algorithm has converged to a local minimum. To quantify the quality of the given solution, we calculate the error with respect to the measurement constraint
\begin{equation}
    \mathrm{error}_M = \sqrt{\sum|\psi - P_M(\psi)|^2/\sum|I|^2},
    \label{eq:error}
\end{equation}
which is the RMS difference between a guess $\psi$ and its projection onto the measurement $P_M(\psi)$, normalized to the RMS value of the measured data $I$. We use the same normalization for the RMS change of the guess.

\begin{figure}[thb!]
\includegraphics[width=0.9\linewidth,center]{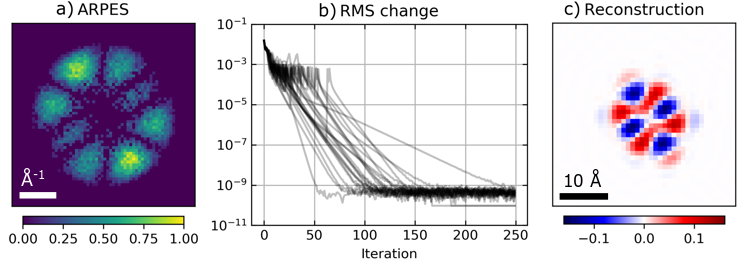}
\caption{\label{fig:sparse50_AP} 
Reconstruction results with $S=50$ allowed non-zero pixels. 
(a) Simulated noisy {photoelectron} momentum map of the HOMO of Coronene. The Poisson noise corresponds to a total electron count of $10^4$.
(b) RMS change per iteration for 25 randomly-initialized orbital reconstructions. 
(c) Averaged result of 25 reconstruction runs. The averaging procedure is described in section~\ref{sec:averaging_procedure}.}
\end{figure}
In order to investigate the efficiency and accuracy of various phase retrieval approaches, we have simulated {photoelectron momentum map data} for the highest occupied molecular orbital (HOMO) of a Coronene molecule in free space as calculated using the ORCA quantum chemistry package \cite{neese_orca_2012}. To characterize the reconstruction accuracy independent of the molecular orbital shape, we have also simulated {momentum map} data for the second highest occupied orbital (HOMO-1), which {is degenerate} with the HOMO. The orbitals appear similar to those presented by L\"uder \emph{et al.} \cite{luder_understanding_2017}. {It should be noted that an experimental analysis of Coronene orbital imaging data is complicated by the degeneracy of HOMO and HOMO-1 and the rotational symmetry of the Coronene molecule, which} means that multiple orientations of the molecular orbital can occur simultaneously \cite{udhardt_influence_2017}. This leads to a {momentum map} in which the rotated contributions {of both states} are overlapped \cite{wiesner_different_2012}, which breaks the direct Fourier transform relationship between measurement and orbital \cite{stadtmuller_orbital_2012}. {In the simulation, we avoid this problem by using a single orbital in one orientation.} The spectra were simulated for photoelectrons with 14.7~eV kinetic energy and an energy spread of 150~meV. The angle of incidence was set to 22\textdegree. Secondary electron yield as well as possible substrate band structure were not included in the simulation, leading to a background-free measurement. 
We have added Poisson noise to the simulated {photoelectron  momentum map} patterns corresponding to a total of $10^4$ photoelectrons. The simulated {momentum maps} are then divided by the known polarization factor to yield the measured momentum map of the orbital. As an example, the momentum map corresponding to the HOMO is shown in Fig.~\ref{fig:sparse50_AP}a. 

\subsection{Initial results}
The results of a series of orbital reconstructions for the HOMO are shown in Fig.~\ref{fig:sparse50_AP}. {Note that the photoelectron momentum map and the real space wavefunction of the coronene HOMO orbital look very similar. For clarity in this paper, therefore, all momentum maps and real space wavefunctions are plotted with a dark and white background, respectively.} The reconstruction series consists here of 25 runs of the alternating projections algorithm, each initialized by a random guess of the orbital. For all these reconstructions we used the same input data (Fig.~\ref{fig:sparse50_AP}a), and the sparsity constraint $S$ was set to 50. After the phase retrieval procedure, we interpolate the retrieved orbital by a factor 2 in both the x and y direction for visualization purposes only. The RMS change on iteration shown in Fig.~\ref{fig:sparse50_AP}b shows that the algorithm converges well in 250 iterations. However, it is not the case that all runs converge to the same solution, even when degrees of freedom such as the sign and position of the orbital are corrected for. This shows that alternating projections converges to local minima. 

In order to select the accurate reconstructions from the individual converged reconstructions, we calculate the reconstruction error given by Eq.~\ref{eq:error}. It is necessary to fix the position and sign of the reconstructions before averaging, as these are not constrained by the phase retrieval procedure. \label{sec:averaging_procedure} The reconstruction with the lowest error is chosen as an example for this procedure. Reconstructions which have an RMS difference with the example larger than 0.3 are removed from the analysis. Finally, the reconstructions are averaged. For the result shown in Fig.~\ref{fig:sparse50_AP}c, 18 out of 25 reconstructions were included in the final result. 

In order to increase the reconstruction chance of the phase retrieval procedure, we have replaced AP with the \emph{relaxed averaged alternating reflections} (RAAR) algorithm \cite{luke_relaxed_2004}. A single iteration of this algorithm is given by
\begin{equation}
    \psi_{n+1} = \frac{1}{2}\beta (R_{S,real} R_M + 1)\psi - (1-\beta) P_M \psi,
\end{equation}
where the reflection operator $R_i$ is defined as $2P_i - 1$ and $\beta$ a damping parameter usually chosen between 0 and 1. This parameter controls the behaviour of the algorithm such that it searches for global minima when $\beta$ is close to one, while smaller values enable local minimization \cite{marchesini_invited_2007}. We have found that smoothly ramping $\beta$ from 0.85 in the beginning to 0.5 after 35 iterations leads to an efficient phase retrieval algorithm. Using otherwise identical settings to those used in Fig.~\ref{fig:sparse50_AP}, we reconstructed a good object in 25 out of 25 runs of the algorithm, with a final reconstruction which is highly similar to the one shown in Fig.~\ref{fig:sparse50_AP}c.

\subsection{Dependence on the sparsity constraint}
\begin{figure}[htb!]
\includegraphics[width=0.9\linewidth,center]{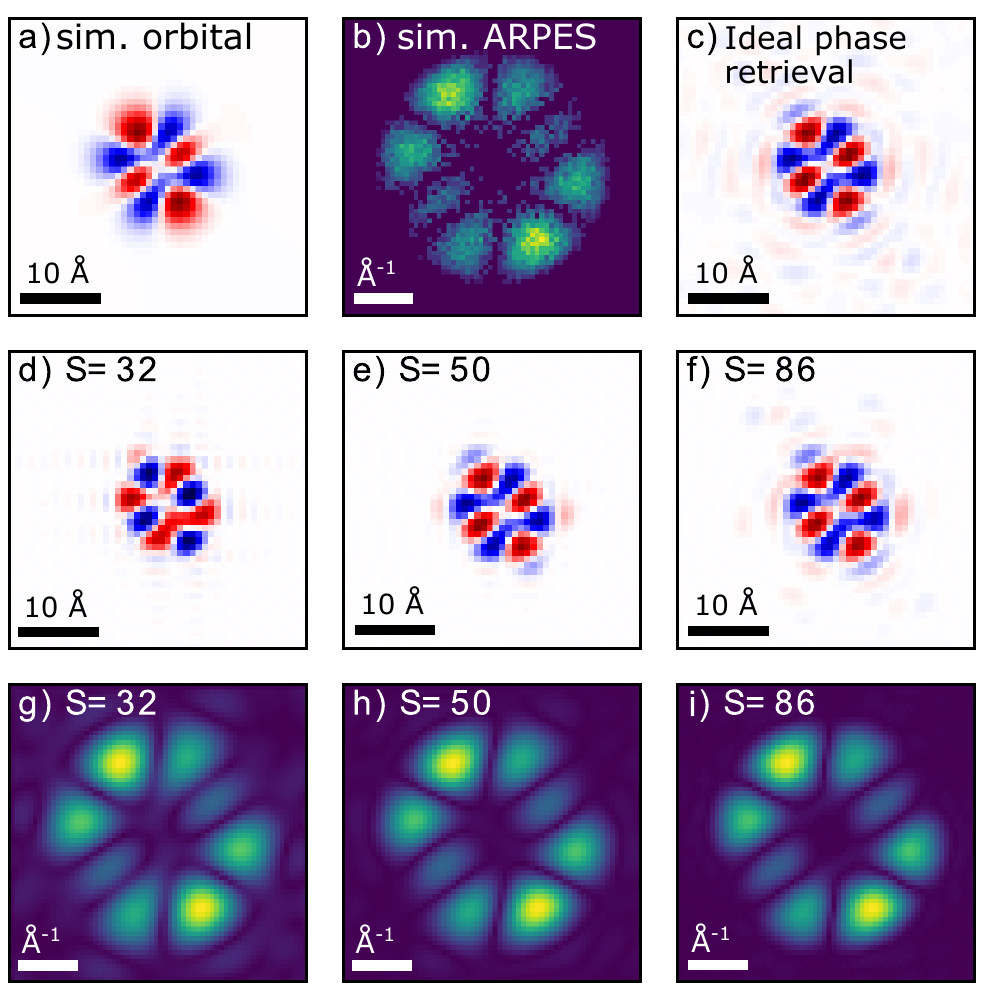}
\caption{\label{fig:3x3_cor_raar} 
Reconstruction results for various sparsity constraints using the RAAR algorithm. 
(a) Value of the simulated HOMO of Coronene at constant height of {1.25~\AA} above the molecular plane. 
(b) Simulated {photoelectron} momentum map for the HOMO of isolated Coronene, assuming $10^4$ detected photoelectrons.
(c) Orbital image resulting from the measured amplitudes and the phase from (a).
(d-f) Average reconstructed orbitals for varying sparsity constraints \textit{S}.
(g-i) Predicted {photoelectron momentum map} calculated by Fourier transform from (d-f) respectively.}
\end{figure}
Although the sparsity number of the to be reconstructed orbital can often be estimated based on prior information, this estimate may not be accurate. In order to investigate how the accuracy of the reconstruction depends on the exact sparsity constraint, we have performed phase retrieval for a wide range of sparsity constraints. These analyses were performed as described in section \ref{sec:averaging_procedure}. Fig.~\ref{fig:3x3_cor_raar} shows the phase retrieval results for three different sparsity constraints (32, 50 and 86), as well as the original calculated HOMO and the simulated {photoelectron momentum map}. For the small sparsity constraint ($S=32$, Fig.~\ref{fig:3x3_cor_raar}d), we retrieve a compact, recognizable orbital, but the symmetry of the orbital is partially broken. Furthermore, the predicted {momentum map} shows non-zero amplitudes at the edges of the spectrum, which leads to ringing in the interpolated orbital. The medium sparsity constraint ($S=50$, Fig.~\ref{fig:3x3_cor_raar}e) leads to the best estimate of the molecular orbital, as compared to the DFT calculation (Fig.~\ref{fig:3x3_cor_raar}a). From the predicted {momentum map}, it is clear that the orbital reconstruction largely corrects for the Poisson noise in the measured data. Nevertheless, the smooth edges of the true orbital are not reproduced. We attribute this to the nature of the ARPES measurement, which measures a hemisphere in the three-dimensional momentum distribution of the orbital. It is therefore also visible in the orbital image that would be retrieved with the true phase calculated from the original orbital. {These artefacts can be reduced through the use of a higher photon energy for photoemission in the future, which increases the size of the photoemission horizon and allows for a higher resolution.} Finally, the largest sparsity constraint ($S=86$, Fig.~\ref{fig:3x3_cor_raar}f) yields a similarly good image surrounded by weak noise. This indicates that the measurement noise is less suppressed by the phase retrieval algorithm.  

\begin{figure}[htb!]
\includegraphics[width=0.9\linewidth,center]{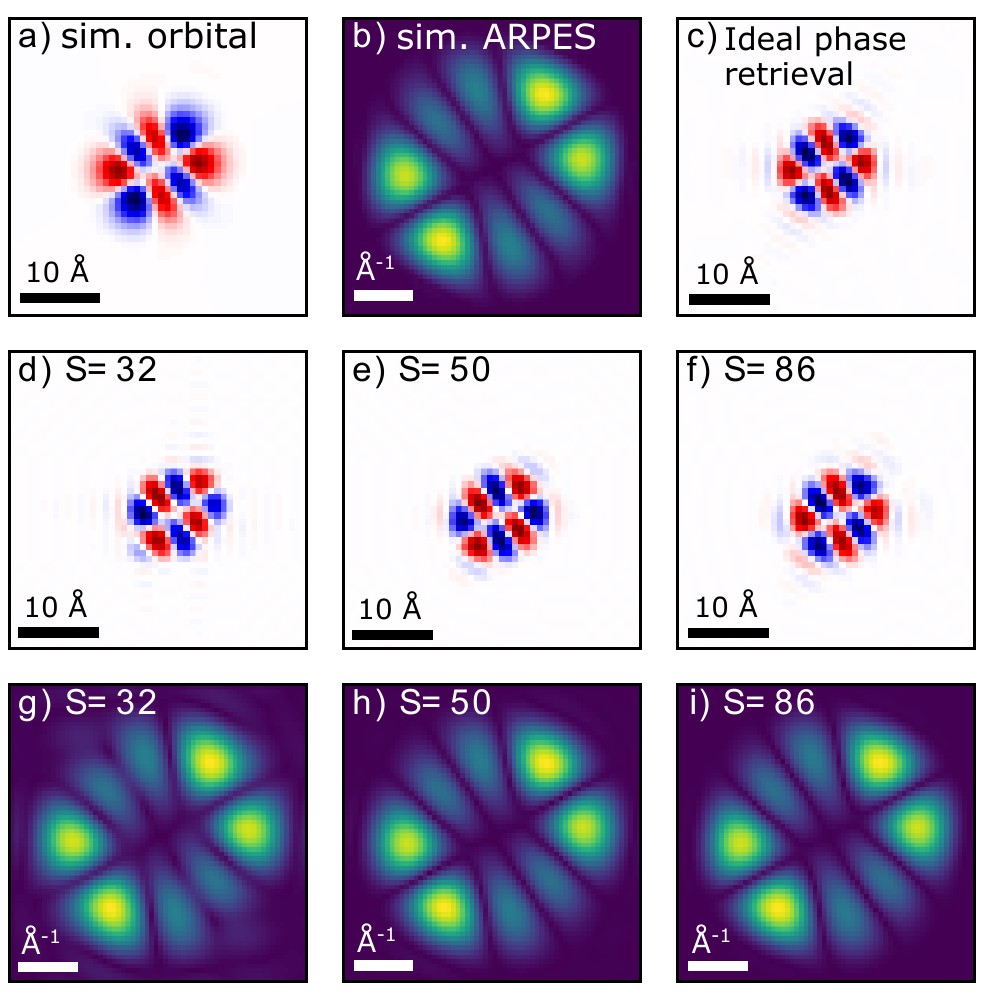}
\caption{\label{fig:3x3_cor2_raar_noisefree} 
Reconstruction results for various sparsity constraints using the RAAR algorithm in the absence of significant noise. 
(a) Value of the simulated HOMO-1 of Coronene at constant height of {1.25~\AA} above the molecular plane. 
(b) Simulated photoelectron  {momentum map} of the HOMO-1 for isolated Coronene, assuming $10^7$ detected photoelectrons.
(c) Orbital image resulting from the measured amplitudes and the phase from (a).
(d-f) Average reconstructed orbitals for varying sparsity constraints \textit{S}.
(g-i) Predicted {momentum maps} calculated by Fourier transform from (d-f) respectively.}
\end{figure}
We have performed an identical analysis for a high-quality simulated {photoelectron momentum map} of the HOMO-1 of isolated Coronene, shown in Fig.~\ref{fig:3x3_cor2_raar_noisefree}. We have assumed a total of $10^7$ counted photoelectrons, leading to much less pronounced Poisson noise. This leads to a significant improvement of the orbital reconstruction using a larger sparsity constraint (Fig.~\ref{fig:3x3_cor2_raar_noisefree}f). From this reconstruction, it is also clear that a larger sparsity constraint does not enable better reconstruction of the smooth edges of the orbital.

\begin{figure}[htb!]
\includegraphics[width=0.8\linewidth,center]{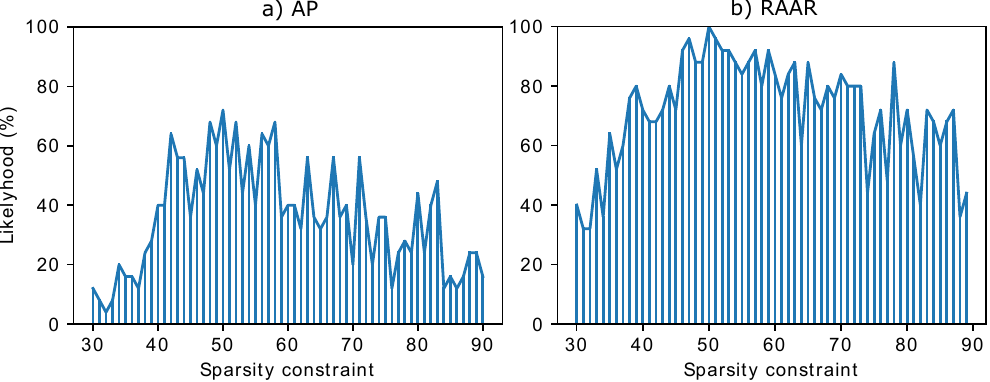}
\caption{
\label{fig:chances_APvsRAAR}
Chance of successful convergence for a single phase retrieval run as function of the sparsity constraint using the alternating projections algorithm (a) or the RAAR algorithm (b), both for the noisy HOMO {photoelectron momentum map} data set.
}
\end{figure}

From Figures~\ref{fig:3x3_cor_raar} and \ref{fig:3x3_cor2_raar_noisefree}, it is evident that the quality of the reconstruction provides an indication of the accuracy of the sparsity constraint. Furthermore, it is clear that an inaccurate sparsity constraint does not necessarily prevent phase retrieval. In Fig.~\ref{fig:chances_APvsRAAR}, the chance that the reconstruction algorithm yields a good solution for a given sparsity constraint is shown. This is calculated from the percentage of solutions included in the final averaged reconstruction using the method described in section~\ref{sec:averaging_procedure}. These results demonstrate that RAAR converges much more successfully to the global minimum than alternating projections. Furthermore, the reconstruction chance shows a clear optimum at $S \approx 50$ for both algorithms. The reconstruction chance diminishes sharply for lower sparsity constraints and more slowly for larger sparsity constraints. The results also show clearly that only a very rough estimate of the sparsity is necessary to achieve an initial orbital reconstruction.

\subsection{Variations}
In the previous analyses, we have limited the reconstructed orbital to real values. This strong constraint is incorporated in step 1 of the sparsity projection. The realness constraint is however not necessary for phase retrieval. In an analysis similar to Fig.~\ref{fig:chances_APvsRAAR}, we have found that RAAR with a complex sparsity constraint has a reconstruction chance which varies between 20\% and 50\% for sparsity parameters between 30 and 90. Nevertheless, good reconstructions can be selected from this procedure based on the measurement error (Eq.~\ref{eq:error})  . 

We have also considered the combination of sparsity with other constraints. A symmetry constraint is particularly attractive due to the high symmetry of the HOMO and HOMO-1 in Coronene. Such a constraint can be implemented by averaging the orbital guess with its mirror image before applying the sparsity constraint. Anti-symmetry can be accounted for by inverting the mirror image before averaging, and the procedure can be repeated for multiple symmetry axes. However, we have found that a combined sparsity and symmetry constraint causes the algorithm to converge to local minima in which the orbital consists of two separated parts. This behaviour can be corrected by also applying a support constraint. For the simulated {photoelectron momentum map} data, this combined support-symmetry-sparsity constraint did not yield a significant improvement over a simple sparsity constraint.

\begin{figure*}[htbp!]
\includegraphics[width=\textwidth,right]{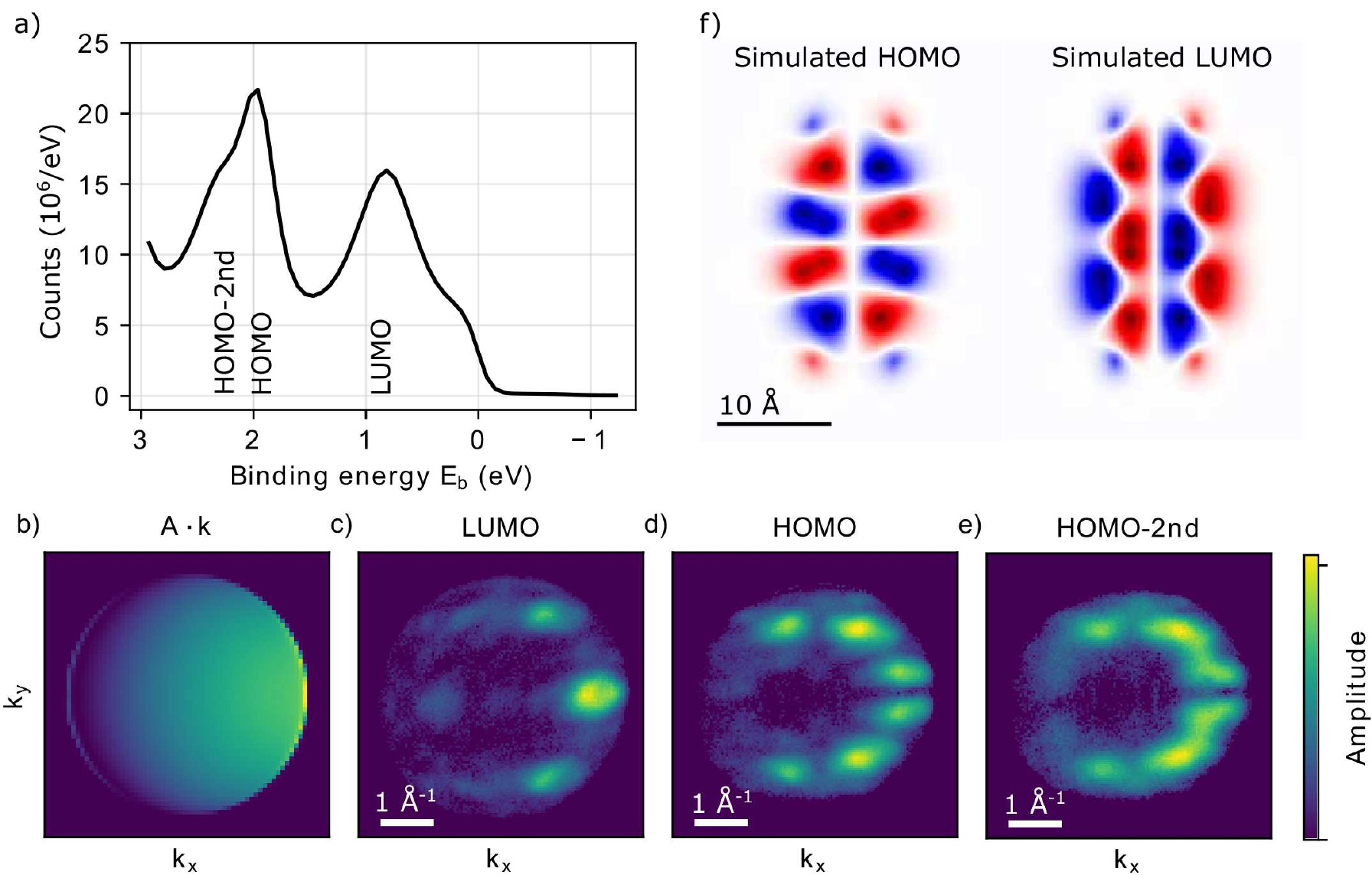}
\caption{
\label{fig:data_overview} 
Overview of the {photoemission momentum microscopy} measurement of PTCDA/Ag(110).
\textbf{a:} Angle-integrated view of the photoelectron spectrum obtained after 2 hours of measurement. 
\textbf{b:} Typical photoemission transfer function $|A \cdot k|$ for our system with a 22\textdegree{} angle of incidence on the sample. 
\textbf{c, d, e:} Photoelectron momentum maps (by amplitude) of the occupied LUMO of the first monolayer, HOMO of the first monolayer and HOMO of the second monolayer respectively. These momentum maps were retrieved by fitting a Gaussian with fixed position and width to the photoemission spectrum measured at individual pixels. 
{\textbf{f:} Simulated HOMO and LUMO orbital of PTCDA in free space as calculated using the ORCA quantum chemistry package \cite{neese_orca_2012}.}
}
\end{figure*} 

\section{Application to experimental data}

\subsection{Experimental details}
We have applied sparsity-driven orbital reconstruction to photoemission spectra of pery\-lene-3,4,\-9,10-te\-tra\-carboxylic di\-anhydride (PTCDA) adsorbed on the Ag(110) surface, prepared by sublimation from a Knudsen cell. We used low-energy electron diffraction (LEED) to confirm that PTCDA is deposited in a brick wall monolayer, where all molecules are aligned along the $[001]$ direction \cite{willenbockel_energy_2013}. 

Fig.~\ref{fig:data_overview}a shows photoelectron spectra near the Fermi edge acquired by integrating the acquired data set over the full photoemission horizon and over a small selected window where a strong molecular signal is seen. In total, we observe three signals that can be attributed to the PTCDA molecule, at binding energies of 0.85~eV, 2.0~eV and 2.3~eV. Of these, we attribute the first two to the LUMO and HOMO of the PTCDA brick wall monolayer, while the third signal is assumed to arise from the HOMO of a partial {PTCDA bilayer} in the herringbone structure \cite{willenbockel_energy_2013, wiesner_electronic_2012}. It is therefore referred to as HOMO-2nd. In a procedure which is similar to the data analysis in \cite{grimm_molecular_2018}, we fit the photoelectron spectrum at each position on the detector with a set of three Gaussian peaks and a linearly changing offset. The peak positions and widths are determined from the data shown in Fig.~\ref{fig:data_overview}a. The amplitudes determined by the fitting procedure then yield the photoelectron momentum maps which are shown in Fig.~\ref{fig:data_overview}c-e. This analysis procedure separates the contributions of the states at 2.0~eV and 2.3~eV and also reduces the presence of noise on the momentum map.

\subsection{Analysis of experimental results}
For the analysis of measured {photoelectron momentum map} patterns, which are asymmetric due to the polarization term $\vec{A}\cdot \vec{k}$ (shown in Fig~\ref{fig:data_overview}b), we have implemented {an accordingly modified} measurement constraint
\begin{equation}
    P_M \psi_n = 
    \left(1-
        \frac{|\vec{A}\cdot \vec{k}|^2 |\psi_n|^2 +2\alpha^2}
        {(|\vec{A}\cdot \vec{k}|^2 |\psi_n|^2 + \alpha^2)^{3/2}}
        \left(\frac{|\vec{A}\cdot \vec{k}|^2 |\psi_n|^2}
            {(|\vec{A}\cdot \vec{k}|^2 |\psi_n|^2 + \alpha^2)^{1/2}} - I^{1/2}
        \right)
    \right)\psi_n,
\end{equation}
in which $\alpha$ is a regularization parameter typically chosen between 0.01 and 10 to prevent division by zero when $|\vec{A}\cdot \vec{k}| |\psi_n|$ is zero. This approach avoids errors in the calculation of the momentum map from the measured data \cite{luke2002optical}. Furthermore, the angle of incidence and regularization parameter can be optimized easily in the phase retrieval process. 

As input-data for the orbital reconstruction, we have used the data presented in Fig.~\ref{fig:data_overview}c and d (LUMO and HOMO, respectively), resampled on a 64x64 pixel grid. This provides sufficient oversampling of the momentum maps while also allowing fast calculation of the Fourier transforms. Although the Gaussian fitting procedure removes most of the background signal due to e.g. secondary photoelectrons, a residual background signal remains. This background is removed by applying a threshold at the most common pixel value inside the photoemission horizon. {The goal of the following phase-retrieval procedure is to reconstruct from these momentum maps the wavefunctions of the molecular orbitals in real space. For comparison, calculated molecular orbitals are shown in Fig. Fig.~\ref{fig:data_overview}f. Note, however, that these calculations are carried out for PTCDA molecules in free space.}

\begin{figure}[htb!]
\includegraphics[width=1\linewidth, center]{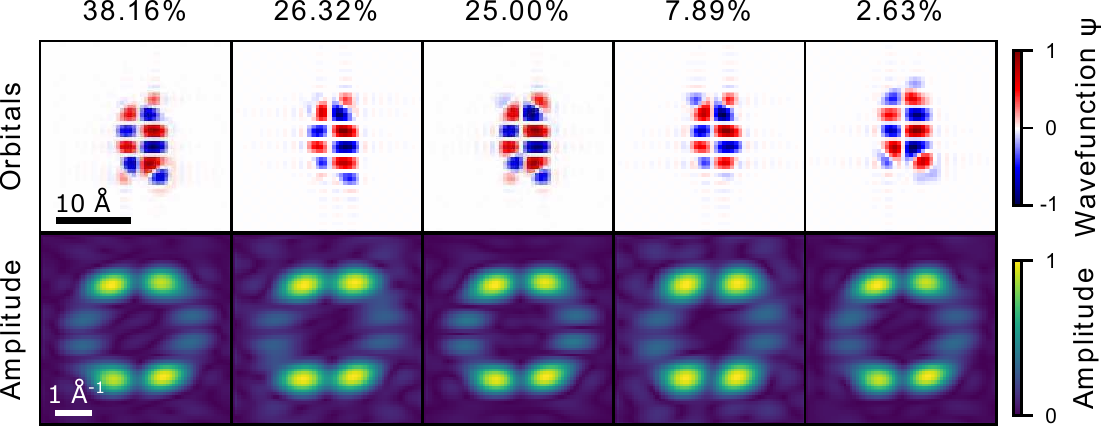}
\caption{
\label{fig:clusters_nonsym} 
Prevalence of typical {reconstructed orbitals for the HOMO of PTCDA using a sparsity constraint between 22 and 38. Note that, despite the measurement constraint is used, the momentum maps show variations in the distributions and do also not exactly match the measured photoelectron momentum distribution. The reason is that in our approach the sparsity constraint is strictly fulfilled, so that the measurement constraint cannot simultaneously be justified in the final result. Therefore, the shown momentum maps can be interpreted as possible noise free solutions corrected for the $|\vec{A}\cdot \vec{k}|$ matrix element.}}
\end{figure} 

For the HOMO {photoelectron momentum map} data as presented in Fig.~\ref{fig:data_overview}d, we have found two phase retrieval approaches which result in recognizable orbitals, one method in which we use only sparsity with a very loose support, and one in which we also enforce the symmetry of the orbital as expected from DFT calculations (see Fig.~\ref{fig:data_overview}f and Ref. \cite{luftner_imaging_2014}). The support we use is calculated by thresholding the absolute autocorrelation of the orbital at 5\% of its maximum value and then dilating the resulting mask by one pixel. This results in a loose support which is roughly twice the size of the orbital. In the absence of a symmetry constraint, we find orbitals which are similar to those presented in Fig.~\ref{fig:data_overview}f using sparsity constraints in the range between 22 and 38 pixels. When the symmetry constraint is included, we find that a wider range of sparsity constraints can be used, with 25 to 100 pixels yielding good results. 

In comparison to the simulated data for the coronene molecule, we observe significantly more variance in the retrieved orbital, as well as a stronger dependence on the settings of the reconstruction algorithm. For a given set of reconstruction parameters, we typically observe a small number of frequently-occurring orbital shapes. These can be ascribed to the local minima in the phase-retrieval landscape. To determine these local minima, we repeat the phase retrieval process with different random initialization while also considering several sparsity constraints. We then use the K-means clustering algorithm to collect and average similar reconstructions which belong to the same local minimum \cite{lloyd_least_1982, arthur_k-means++:_2007}. Here, it should be noted that an orbital reconstruction may be shifted, mirrored or sign-flipped without change in the predicted momentum distribution. We therefore correct for these degrees of freedom before clustering by minimizing the RMS difference of the orbital guess with an example guess.

\begin{figure}[htb!]
\includegraphics[width=1\linewidth,center]{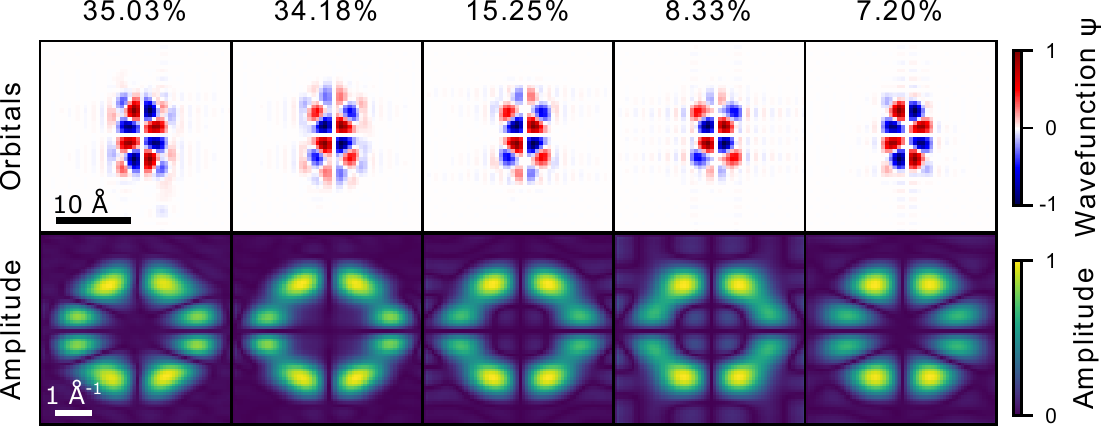}
\caption{
\label{fig:clusters_sym} 
Prevalence of typical reconstructed orbitals for the HOMO of PTCDA and their predicted momentum distributions reconstructed using a sparsity constraint between 25 and 100 in combination with a symmetry constraint.
}
\end{figure} 

An overview of the results for the symmetry-free orbital reconstruction of the HOMO is shown in Fig.~\ref{fig:clusters_nonsym}. A similar overview for the symmetry-con\-strained re\-con\-struc\-tion of the HOMO is shown in Fig.~\ref{fig:clusters_sym}. In both cases, we have sorted the orbital reconstructions in five clusters of similar reconstructions using the K-means algorithm. The figures then show the average orbital of each cluster, the likelihood of a given phase retrieval run resulting in that orbital and the Fourier transform of the orbital, from which the {photoelectron momentum map} pattern may be calculated. 

Out of all reconstructions, the first symmetry-constrained reconstruction best approaches the density functional theory (DFT) calculation from Fig.~\ref{fig:data_overview}f. Nevertheless, all reconstructions match roughly in size and shape with the shape of the PTCDA molecule. For the symmetry-free reconstructions, {the amplitude of the lobes is allowed to be asymmetrically distributed, which is also seen in the reconstructions (Fig.~\ref{fig:clusters_nonsym}). The more pronounced central eight lobes are reconstructed well, but their shape is more rectangular than expected. The less pronounced corner lobes show a strong asymmetry  - often, one or two corner lobes are missing. Here, the symmetry-constrained reconstruction is clearly advantageous, where the amplitude of the corner lobes is enforced to be evenly distributed.}

For the symmetry-constrained reconstructions, we observe two classes of re\-con\-struc\-tions which occur roughly equally often: those with zeros on the diagonal in the momentum distribution, and those where those zeroes do not occur. Apart from the true orbital, there are a number of reasons which may cause a lack of zeroes along these diagonals, such as blurring due to aberration of the momentum microscope, a residual background, a residual contribution from the HOMO of the second monolayer or possibly a bad centering of the momentum map. It is therefore not trivial to indicate which of these orbitals best describes the PTCDA/Ag(110) orbital based only on the orbital imaging results. Here, we observe that the measurement suggests a more rectangular shape of the orbital and the individual lobes, and the more triangular shape suggested by DFT is only reproduced if the strong symmetry constraint is used. 

\begin{figure}[htb!]
\includegraphics[width=1\linewidth,center]{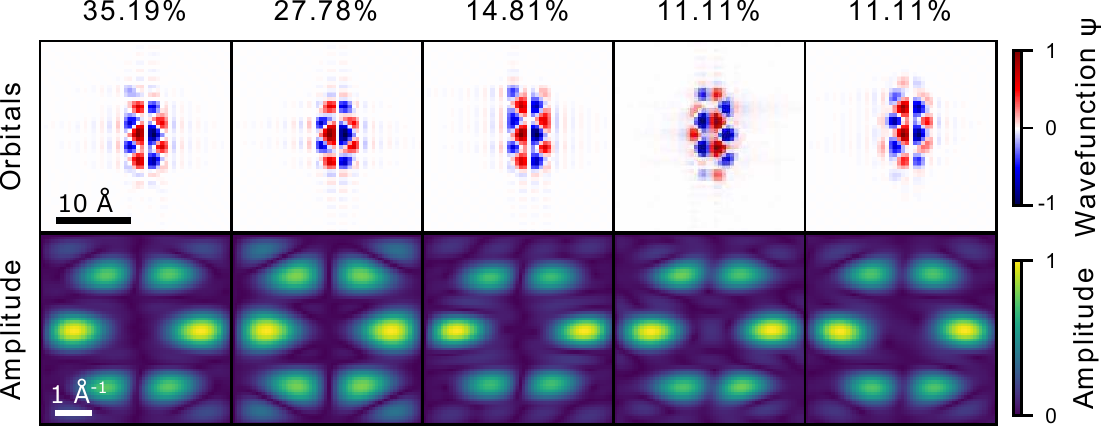}
\caption{
\label{fig:clusters_nonsym_lumo} 
Prevalence of typical reconstructed orbitals for the LUMO of PTCDA and their predicted momentum distributions reconstructed using a sparsity constraint between 12 and 20. 
}
\end{figure} 

\begin{figure}[htb!  ]
\includegraphics[width=1\linewidth,center]{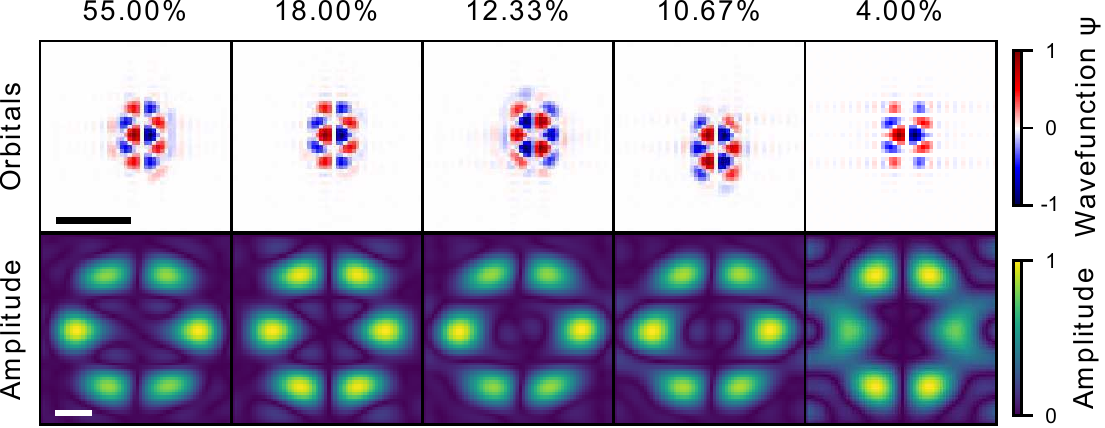}
\caption{
\label{fig:clusters_sym_lumo} 
Prevalence of typical orbital guesses for the LUMO of PTCDA and their predicted momentum distributions reconstructed using a sparsity constraint between 25 and 100 in combination with a symmetry constraint.
}
\end{figure}

As mentioned before, a crucial advantage of the momentum microscope is that it enables the measurement of multiple molecular orbitals simultaneously. In order to demonstrate this capability, we have performed the same analysis to the momentum distribution acquired for the LUMO, the results of which are shown in Figures~\ref{fig:clusters_nonsym_lumo} and \ref{fig:clusters_sym_lumo}. The quality of these reconstructions is similar as for the HOMO, and again, the phase retrieval process finds a number of different orbital shapes. A crucial difference here is that the triangular shape of the lobes suggested by DFT {(see Fig.~\ref{fig:data_overview}f)} is reproduced in the measurement also without a symmetry constraint. Figures~\ref{fig:clusters_nonsym_lumo} and \ref{fig:clusters_sym_lumo} also show that the phase retrieval algorithm can lead to orbitals which have a non-zero momentum distribution outside of the photoemission horizon. This produces sharp features in the reconstructed orbital which the true orbital does not necessarily have. It may therefore be better to set these high-momentum components to zero.

\section{Concluding remarks}
In summary, we {present first orbital imaging data with our newly developed femtosecond momentum microscopy setup \cite{keunecke_time-resolved_2020}. Using a combination of time-of-flight-based photoemission and a high-repetitive high-harmonic generation light source, we are able to collect energy and momentum resolved photoemission data without the need for scanning either energy or momentum direction. This approach makes this setup very efficient and ideally suited for orbital imaging experiments. Moreover, it opens the door to time-resolved orbital imaging experiments in the future. We also present a significant step forward it the orbital imaging data analysis procedure, and have} investigated the use of a sparsity constraint for the reconstruction of molecular orbitals. In com\-par\-i\-son to the commonly used support constraint, a sparsity constraint is easy to determine, which greatly simplifies the image reconstruction. We find that a combined sparsity and realness constraint, which is most relevant in the context of orbital imaging, enables accurate and efficient reconstruction of molecular orbitals even in the presence of significant Poisson noise. Although it is clear that the value of the sparsity constraint influences the quality of the reconstruction, good reconstructions can be retrieved for a wide range of sparsity parameters. 

We have applied the developed sparsity-driven orbital reconstruction to ex\-per\-i\-men\-tal {photoelectron momentum map} data of a PTCDA monolayer on the silver (110) surface. The measurement data were acquired using time-of-flight-based momentum microscope, which enabled the simultaneous measurement of the HOMO and LUMO of PTCDA. As for the simulations, we find good reconstructions for a broad range of sparsity values. We note however, that the phase retrieval algorithm does not always result in the same orbital, and that the predicted momentum distribution also varies. It is therefore necessary to repeat the phase retrieval procedure, and to select the most likely orbital shape based on the the results of this procedure, possible prior knowledge of the orbital and a good understanding of the limitations of the measurement. 

Finally, we note that the sparsity-driven reconstruction procedure should also be applicable in the three-dimensional case. For future experiments, it is promising to apply sparsity-based methods to full three-dimensional orbital tomography \cite{weis_exploring_2015}.

\section*{Acknowledgments}
We would like to thank Benjamin Stadtm\"uller for helpful discussions. {G.S.M.J. acknowledges support from the Alexander von Humboldt Foundation.} M.K., C.M., D.S. and S.M. acknowledge support from the German Science Foundation through SFB 1073, project B07. D.R.L. acknowledges support from the German Science Foundation through RTG2088. S.S. acknowledges the Dorothea Schlözer Postdoctoral Program for Women.

The phase retrieval was performed using the python package \emph{ProxToolbox}, which is available from http://num.math.uni-goettingen.de/proxtoolbox.

\section*{References}
\bibliographystyle{iopart-num}
\bibliography{paper_sparse_orbital}

\providecommand{\newblock}{}
\begin{thebibliography}{10}
\expandafter\ifx\csname url\endcsname\relax
  \def\url#1{{\tt #1}}\fi
\expandafter\ifx\csname urlprefix\endcsname\relax\def\urlprefix{URL }\fi
\providecommand{\eprint}[2][]{\url{#2}}

\bibitem{puschnig_reconstruction_2009}
Puschnig P, Berkebile S, Fleming A~J, Koller G, Emtsev K, Seyller T, Riley J~D,
  Ambrosch-Draxl C, Netzer F~P and Ramsey M~G 2009 {\em Science\/} {\bf 326}
  702--706 ISSN 0036-8075, 1095-9203

\bibitem{luftner_imaging_2014}
L\"uftner D, Ules T, Reinisch E~M, Koller G, Soubatch S, Tautz F~S, Ramsey M~G
  and Puschnig P 2014 {\em Proceedings of the National Academy of Sciences\/}
  {\bf 111} 605--610 ISSN 0027-8424, 1091-6490

\bibitem{weis_exploring_2015}
Wei{\ss} S, L\"uftner D, Ules T, Reinisch E~M, Kaser H, Gottwald A, Richter M,
  Soubatch S, Koller G, Ramsey M~G, Tautz F~S and Puschnig P 2015 {\em Nature
  Communications\/} {\bf 6} 8287 ISSN 2041-1723

\bibitem{willenbockel_energy_2013}
Willenbockel M, Stadtmüller B, Sch\"onauer K, Bocquet F~C, L\"uftner D,
  Reinisch E~M, Ules T, Koller G, Kumpf C, Soubatch S, Puschnig P, Ramsey M~G
  and Tautz F~S 2013 {\em New Journal of Physics\/} {\bf 15} 033017 ISSN
  1367-2630

\bibitem{kliuiev_application_2016}
Kliuiev P, Latychevskaia T, Osterwalder J, Hengsberger M and Castiglioni L 2016
  {\em New Journal of Physics\/} {\bf 18} 093041 ISSN 1367-2630

\bibitem{tautz_structure_2007}
Tautz F~S 2007 {\em Progress in Surface Science\/} {\bf 82} 479--520 ISSN
  0079-6816

\bibitem{schwalb_electron_2008}
Schwalb C~H, Sachs S, Marks M, Schöll A, Reinert F, Umbach E and Höfer U 2008
  {\em Physical Review Letters\/} {\bf 101} 146801

\bibitem{hagen_electronic_2010}
Hagen S, Luo Y, Haag R, Wolf M and Tegeder P 2010 {\em New Journal of
  Physics\/} {\bf 12} 125022 ISSN 1367-2630

\bibitem{varene_ultrafast_2012}
Varene E, Bogner L, Bronner C and Tegeder P 2012 {\em Physical Review
  Letters\/} {\bf 109} 207601

\bibitem{steil_spin-dependent_2013}
Steil S, Großmann N, Laux M, Ruffing A, Steil D, Wiesenmayer M, Mathias S,
  Monti O~L~A, Cinchetti M and Aeschlimann M 2013 {\em Nature Physics\/} {\bf
  9} 242--247 ISSN 1745-2481

\bibitem{marks_formation_2014}
Marks M, Schöll A and Höfer U 2014 {\em Journal of Electron Spectroscopy and
  Related Phenomena\/} {\bf 195} 263--271 ISSN 0368-2048

\bibitem{schonauer_charge_2016}
Schönauer K, Weiss S, Feyer V, Lüftner D, Stadtmüller B, Schwarz D, Sueyoshi
  T, Kumpf C, Puschnig P, Ramsey M~G, Tautz F~S and Soubatch S 2016 {\em
  Physical Review B\/} {\bf 94} 205144

\bibitem{miao_extending_1999}
Miao J, Charalambous P, Kirz J and Sayre D 1999 {\em Nature\/} {\bf 400}
  342--344 ISSN 0028-0836

\bibitem{fienup_reconstruction_1987}
Fienup J~R 1987 {\em JOSA A\/} {\bf 4} 118--123 ISSN 1520-8532

\bibitem{marchesini_invited_2007}
Marchesini S 2007 {\em Review of Scientific Instruments\/} {\bf 78} 011301 ISSN
  0034-6748

\bibitem{sayre_implications_1952}
Sayre D 1952 {\em Acta Crystallographica\/} {\bf 5} 843--843 ISSN 0365-110X

\bibitem{fienup_reconstruction_1982}
Fienup J~R, Crimmins T~R and Holsztynski W 1982 {\em JOSA\/} {\bf 72} 610--624

\bibitem{marchesini_x-ray_2003}
Marchesini S, He H, Chapman H~N, Hau-Riege S~P, Noy A, Howells M~R, Weierstall
  U and Spence J~C~H 2003 {\em Physical Review B\/} {\bf 68} 140101

\bibitem{kliuiev_algorithms_2018}
Kliuiev P, Latychevskaia T, Zamborlini G, Jugovac M, Metzger C, Grimm M,
  Schöll A, Osterwalder J, Hengsberger M and Castiglioni L 2018 {\em Physical
  Review B\/} {\bf 98} 085426

\bibitem{elser_phase_2003}
Elser V 2003 {\em JOSA A\/} {\bf 20} 40--55 ISSN 1520-8532

\bibitem{jaganathan_sparse_2013}
Jaganathan K, Oymak S and Hassibi B 2013 Sparse phase retrieval: {Convex}
  algorithms and limitations {\em 2013 {IEEE} {International} {Symposium} on
  {Information} {Theory}\/} pp 1022--1026

\bibitem{shechtman_gespar:_2014}
Shechtman Y, Beck A and Eldar Y~C 2014 {\em IEEE Transactions on Signal
  Processing\/} {\bf 62} 928--938 ISSN 1053-587X

\bibitem{yokoyama_sparse_2019}
Yokoyama Y, Arima T~h, Okada M and Yamasaki Y 2019 {\em J. Phys. Soc. Jpn.\/}
  {\bf 88} 024009 ISSN 0031-9015 publisher: The Physical Society of Japan

\bibitem{medjanik_direct_2017}
Medjanik K, Fedchenko O, Chernov S, Kutnyakhov D, Ellguth M, Oelsner A,
  Schönhense B, Peixoto T~R~F, Lutz P, Min C~H, Reinert F, Däster S, Acremann
  Y, Viefhaus J, Wurth W, Elmers H~J and Schönhense G 2017 {\em Nature
  Materials\/} {\bf 16} 615--621 ISSN 1476-4660

\bibitem{passlack_space_2006}
Passlack S, Mathias S, Andreyev O, Mittnacht D, Aeschlimann M and Bauer M 2006
  {\em Journal of Applied Physics\/} {\bf 100} 024912 ISSN 0021-8979

\bibitem{hellmann_vacuum_2009}
Hellmann S, Rossnagel K, Marczynski-Bühlow M and Kipp L 2009 {\em Phys. Rev.
  B\/} {\bf 79} 035402

\bibitem{schonhense_multidimensional_2018}
Schönhense B, Medjanik K, Fedchenko O, Chernov S, Ellguth M, Vasilyev D,
  Oelsner A, Viefhaus J, Kutnyakhov D, Wurth W, Elmers H~J and Schönhense G
  2018 {\em New J. Phys.\/} {\bf 20} 033004 ISSN 1367-2630

\bibitem{keunecke_time-resolved_2020}
Keunecke M, Möller C, Schmitt D, Nolte H, Jansen G~S~M, Reutzel M, Gutberlet
  M, Halasi G, Steil D, Steil S and Mathias S 2020 {\em arXiv:2003.01602
  [physics]\/} ArXiv: 2003.01602

\bibitem{neese_orca_2012}
Neese F 2012 {\em Wiley Interdisciplinary Reviews: Computational Molecular
  Science\/} {\bf 2} 73--78 ISSN 1759-0884

\bibitem{luder_understanding_2017}
L\"uder J, Ho Cheow M and Manzhos S 2017 {\em Physical Chemistry Chemical
  Physics\/} {\bf 19} 13195--13209

\bibitem{udhardt_influence_2017}
Udhardt C, Otto F, Kern C, L\"uftner D, Huempfner T, Kirchhuebel T, Sojka F,
  Meissner M, Schr\"oter B, Forker R, Puschnig P and Fritz T 2017 {\em The
  Journal of Physical Chemistry. C, Nanomaterials and Interfaces\/} {\bf 121}
  12285--12293 ISSN 1932-7447

\bibitem{wiesner_different_2012}
Wießner M, Lastra N~S~R, Ziroff J, Forster F, Puschnig P, Dössel L, Müllen
  K, Schöll A and Reinert F 2012 {\em New J. Phys.\/} {\bf 14} 113008 ISSN
  1367-2630

\bibitem{stadtmuller_orbital_2012}
Stadtmüller B, Willenbockel M, Reinisch E~M, Ules T, Bocquet F~C, Soubatch S,
  Puschnig P, Koller G, Ramsey M~G, Tautz F~S and Kumpf C 2012 {\em EPL
  (Europhysics Letters)\/} {\bf 100} 26008 ISSN 0295-5075

\bibitem{luke_relaxed_2004}
Luke D~R 2004 {\em Inverse Problems\/} {\bf 21} 37--50 ISSN 0266-5611

\bibitem{wiesner_electronic_2012}
Wießner M, Hauschild D, Schöll A, Reinert F, Feyer V, Winkler K and Krömker
  B 2012 {\em Phys. Rev. B\/} {\bf 86} 045417

\bibitem{grimm_molecular_2018}
Grimm M, Metzger C, Graus M, Jugovac M, Zamborlini G, Feyer V, Sch\"oll A and
  Reinert F 2018 {\em Physical Review B\/} {\bf 98} 195412

\bibitem{luke2002optical}
Luke D~R, Burke J~V and Lyon R~G 2002 {\em SIAM Review\/} {\bf 44} 169--224
  ISSN 0036-1445
  \urlprefix\url{https://epubs.siam.org/doi/abs/10.1137/S003614450139075}

\bibitem{lloyd_least_1982}
Lloyd S 1982 {\em IEEE Transactions on Information Theory\/} {\bf 28} 129--137
  ISSN 0018-9448, 1557-9654

\bibitem{arthur_k-means++:_2007}
Arthur D and Vassilvitskii S 2007 k-means++: {The} {Advantages} of {Careful}
  {Seeding} {\em Proceedings of the eighteenth annual {ACM}-{SIAM} symposium on
  {Discrete} algorithms\/} (Society for Industrial and Applied Mathematics) pp
  1027--1035

\end{thebibliography}

\end{document}